\documentclass[doublecol,linenumber]{our}

\usepackage{amsmath}
\usepackage{braket}

\title{Perturbative treatment of inter-site couplings in the local description of open quantum networks}

\author{A. S. Trushechkin \and I. V. Volovich}

\institute{                    
Steklov Mathematical Institute of the Russian Academy of Sciences - Gubkina 8, Moscow 119991, Russia
}

\abstract{
The problem of construction of a quantum master equation for a system of sites weakly coupled to each other and to one or more reservoirs (open quantum network) is considered. Microscopic derivation of a quantum master equation requires a diagonalization of the Hamiltonian of the network, which can be a difficult task. When the inter-site couplings are weak, the local approach, which neglects the influence of the inter-site couplings on the system-reservoir couplings, is often used. Recently, some doubts were cast to the consistency of the local approach. We develop a systematic perturbation expansion to derive corrections to the local approach and establish its range of validity. Using this extension of the local approach, we derive an expression for the heat flux for a particular model and show that it does not violate the second law of thermodynamics.
}

\begin{document}

\maketitle

\section{Introduction}

Quantum master equation framework is commonly used in the description of charge and energy transfer in quantum networks \cite{MayKuhn}. Quantum master equation can be microscopically derived, but this requires a diagonalization of the Hamiltonian of the network. This can be a difficult task. Even if not, this master equation can be complicated and, hence, difficult for analysis. This approach is sometimes called global, because it considers the network as a whole. When the inter-site couplings in the network are weak, it seems reasonable to neglect their influence on the system-reservoir couplings. This approach is sometimes called local and is often used as a simplification, since it does not require a diagonalization of the full network Hamiltonian. 

However, recently several authors \cite{LevyKosloff,Manrique,Decordi} reported the discrepancies between the local and global descriptions of open quantum systems and, in particular, the violation of the second law of thermodynamics by the local approach \cite{LevyKosloff}. 

We show that the viewpoint of perturbation theory allows to justify this approach, establish its range of validity and derive corrections to it. The small parameter is the inter-site coupling constant. In this formalism, we derive an expression for the heat flux for the model analysed in \cite{LevyKosloff} and show that it does not violate the second law of thermodynamics.

\section{Global and local quantum master equations}

Consider a large system consisting of a relevant system (``system'') and a reservoir. The Hamiltonian has the form

\begin{equation}
H=H_S+H_R+\lambda H_I.
\end{equation}
Here $H_S$ and $H_R$ specify the free dynamics of the system and the reservoir accordingly. The interaction is specified by the term
$H_I=\sum_\alpha A_\alpha\otimes B_\alpha,$
where $A_\alpha$ act on the system, $B_\alpha$ act on the reservoir, and $\lambda$ is a small dimensionless parameter. The application of weak coupling limit with respect to $\lambda$ leads to the following master equation for the reduced density matrix $\rho=\rho(t)$ of the system \cite{AccLuVol,BP,Davies}:
\begin{equation}\label{EqMasterGlob}
\frac{\upd\rho}{\upd t}=-i[H_S+\lambda^2 H_{LS},\rho]+\lambda^2 \mathcal D(\rho),
\end{equation}
where
\begin{equation}\label{EqMasterDissip}
\begin{split}
\mathcal D(\rho)=\sum_\omega\sum_{\alpha,\alpha'}\gamma_{\alpha\alpha'}(\omega)
(&A_{\alpha'}(\omega)\rho A_\alpha(\omega)^\dag\\-\frac12\{&A_\alpha(\omega)^\dag A_{\alpha'}(\omega),\rho\})
\end{split}
\end{equation}
is a dissipator and
\begin{equation}\label{EqMasterLS}
H_{LS}=\sum_\omega\sum_{\alpha,\alpha'}S_{\alpha\alpha'}(\omega)
A_\alpha(\omega)^\dag A_{\alpha'}(\omega)
\end{equation}
is a Lamb shift Hamiltonian (commutes with $H_S$). Here $\omega$ are  Bohr frequencies, i.e. differences between eigenenergies of $H_S$. The summation is performed over all such differences (positive and negative). Here $[\cdot,\cdot]$ and $\{\cdot,\cdot\}$ denote a commutator and an anti-commutator correspondingly. Further, for an arbitrary operator $T$,
\begin{equation}\label{EqMasterObkl}
T(\omega)=\sum_{\varepsilon'-\varepsilon=\omega}P(\varepsilon) T  P(\varepsilon'),
\end{equation}
where $\varepsilon$ and $\varepsilon'$ are eigenenergies of $H_S$ (the summation is performed over all such pairs with the  difference $\omega$), $P(\varepsilon)$ and $P(\varepsilon')$ are projectors onto the corresponding eigenspaces. Finally, $\gamma_{\alpha\alpha'}(\omega)$ and $S_{\alpha\alpha'}(\omega)$ are correlation functions of the reservoir (defined by the operators $B_\alpha$ and the steady state of the reservoir); we will not need their exact expressions. If the reservoir is in a thermal state with the inverse temperature $\beta$, the following relation takes place:
\begin{equation}\label{EqGamma}
\gamma_{\alpha\alpha'}(-\omega)=\exp(-\beta\omega)\gamma_{\alpha'\alpha}(\omega).
\end{equation}

Note that the stochastic limit technique \cite{AccLuVol} allows to derive not only a master equation for the reduced dynamics of the relevant system, but a stochastic unitary dynamics of the large system (the relevant system and the reservoir together).

So, a master equation can be microscopically derived, but this requires a diagonalization of the system Hamiltonian. This approach (and the master equation itself) are sometimes called global, because it considers the system as a whole.

But the diagonalization of the system Hamiltonian can be a difficult task. Even if not, the master equation can be complicated and, hence, difficult for analysis. To simplify the task, suppose that the system Hamiltonian has the form

\begin{equation}\label{EqHs}
H_S=H_0+\nu V,
\end{equation}
where the diagonalization of $H_0$ is easy, and a dimensionless constant $\nu$ is small. A particular case which we are interested in is a network of sites weakly coupled to each other and to a reservoir. In this case, $H_0=\sum_j H_j$ is a sum of free Hamiltonians of each site and $V$ specifies their couplings; the diagonalization of $H_0$ is often trivial. 

Of course, this situation is studied in the well-known framework of perturbation theory \cite{LL}. Suppose that we know eigenvalues $\{\varepsilon^{(0)}_n\}$ and the corresponding 
eigenstates $\{\ket{e^{(0)}_n}\}$ of $H_0$. The eigenvalues and eigenstates of $H$ can be represented by the series
\begin{eqnarray}\label{EqPertEn}
\varepsilon_n&=&\varepsilon_n^{(0)}+\nu\varepsilon_n^{(1)}+\nu^2\varepsilon_n^{(2)}+
\ldots,\\\label{EqPertSt}
\ket{e_n}&=&\ket{e_n^{(0)}}+\nu\ket{e_n^{(1)}}+\nu^2\ket{e_n^{(2)}}+
\ldots,
\end{eqnarray}
where the terms can be found explicitly. Suppose that the perturbation $\nu V$ does not change the multiplicities of the zeroth-order eigenenergies (the situation when this is not the case will be discussed later).  Then, as a zeroth-order approximation we can just put $\varepsilon_n=\varepsilon_n^{(0)}+O(\nu)$ and $\ket{e_n}=\ket{e_n^{(0)}}+O(\nu)$. Accordingly, equations (\ref{EqMasterGlob})--(\ref{EqMasterLS}) are changed to
\begin{equation}\label{EqMasterLoc}
\frac{\upd\rho}{\upd t}=-i[H_0+\nu V+\lambda^2 H^{(0)}_{LS},\rho]+
\lambda^2 \mathcal D^{(0)}(\rho)+O(\lambda^2\nu),
\end{equation}
\begin{equation*}
\begin{split}
\mathcal D^{(0)}(\rho)=\sum_{w}
\sum_{\alpha,\alpha'}\gamma_{\alpha\alpha'}(w)
(&A^{(0)}_{\alpha'}(w)\rho A^{(0)}_\alpha(w)^\dag\\
-\frac12\{&A^{(0)}_\alpha(w)^\dag A^{(0)}_{\alpha'}(w),\rho\}),
\end{split}
\end{equation*}
\begin{equation*}
H_{LS}^{(0)}=\sum_{w}\sum_{\alpha,\alpha'}S_{\alpha\alpha'}(w)
A^{(0)}_\alpha(w)^\dag A^{(0)}_{\alpha'}(w),
\end{equation*}
where $w\equiv\omega^{(0)}$  are differences between zeroth-order eigenenergies $\varepsilon^{(0)}\equiv E$, and
$$T^{(0)}(w)=\sum_{E'-E=w}P^{(0)}(E) T  P^{(0)}(E')$$
for an arbitrary operator $T$. Here $P^{(0)}(E)$ are projectors onto the zeroth-order eigenspaces.

This is exactly what is called the local approach to open quantum dynamics: the inter-site couplings does not affect the system-reservoir couplings.

We see that the local approach can be justified on the base of the global one in case when the inter-site couplings are weak. Also note that both $\lambda$ and $\nu$ are small parameters. By this reason, terms of orders $\nu$ and $\lambda^2$ are included  in (\ref{EqMasterLoc}), but terms of order $O(\lambda^2\nu)$ are not since this is a higher order of smallness.

We assert that all discrepancies between the global and local approaches reported by various authors fall into at least one of two situations: either the inter-site couplings are not small or the analysed quantities are of order $O(\lambda^2\nu)$, which is beyond the local approach's range of validity. 

Let us give an example of the second situation. In \cite{LevyKosloff}, the authors consider a simple network with two sites each of which is coupled to its own  reservoir in a thermal state with different temperatures (see the next section), derive the expression for the steady-state heat flux based on the local approach, and show that, for some values of parameters, there is a heat flux from the cold reservoir into the hot one, which violates the second law of thermodynamics. But the steady-state heat flux found in \cite{LevyKosloff} is proportional to $\lambda^2\nu^2$ (in our notations). Since the local equation (\ref{EqMasterLoc}) is written only up to terms of order $\lambda^2$, the heat flux should be set to zero in this order of smallness, and, hence, there is no violation of the second law of thermodynamics! 

If we want to derive a nontrivial expression for the steady-state heat flux of order $\lambda^2\nu^2$ on the base of the local description, we should take the right-hand side of equation (\ref{EqMasterGlob}) with the eigenenergies and their eigenstates expanded up to the second order in  (\ref{EqPertEn})--(\ref{EqPertSt}). We are going to do this in the next section.

Based on their analysis, the authors of \cite{LevyKosloff} conclude that ``the local approach is only valid for local observables such as the population of each node [site], and is not valid for non-local observables describing energy fluxes''. Here we want to precise this statement: local approach is valid for observables of order $\lambda^2$, but not of  order $O(\lambda^2\nu)$. Of course, in the model considered in \cite{LevyKosloff}, where all non-local observables vanishes whenever $\nu=0$, this is equivalent to the cited statement.

Other words, phenomenological consideration of the local approach does not give its range of validity and leads to discrepancies with the microscopically justified global approach and to violation of the second law of thermodynamics. In contrast, the viewpoint of perturbation theory naturally gives the range of validity of this approach and offers a way of obtaining corrections to it.

An interesting question concerns the relation between the small parameters $\lambda$ and $\nu$ (e.g., whether $\lambda\ll\nu\ll1$ or $\nu\ll\lambda\ll1$, etc.). In principle, any relation is possible. But we should take into account that the corrections to the master equation (\ref{EqMasterGlob}) proportional to higher orders of $\lambda$, in principle, can be derived. First steps in this direction have been done in \cite{PechVol} in the framework of stochastic limit mentioned above. If we drop the term of order $O(\lambda^3)$,  the expansion of the eigenenergies and eigenstates in (\ref{EqPertEn})--(\ref{EqPertSt}) up to the $n$th order makes sense only if $\nu^n\ll\lambda$. 

Physically, $\nu$ can be interpreted as the ratio of a characteristic shift of energy levels to the characteristic difference between the non-perturbed levels. Also $\lambda^2$ can be interpreted as a ratio of the time scale of relaxation of the reservoir to the time scale of evolution of the state of the system.

Let us also mention that a perturbative treatment of inter-site couplings in the local description was also considered in \cite{Perturb}. But this was done in a different way, which in some cases (e.g. degenerate eigenenergies or degenerate Bohr frequencies, i.e. if there are different pairs of energies with the same difference) produces indefinitely increasing (with time) terms.

For recent considerations of the degenerate case, see \cite{AKV}.

\section{Two-site model}

Consider a model of two sites $A$ and $B$ each of which is coupled to its own reservoir in a thermal state with different temperatures $T_h$ and $T_c$, accordingly ($T_h>T_c$). The corresponding inverse temperatures are $\beta_h$ and $\beta_c$. The sites are either two two-level systems or two harmonic oscillators. The Hamiltonian is
$$H=H_0+\nu V+H_h+H_c+\lambda H_{Ah}+\lambda H_{Bc},$$
where $H_0=E_A a^\dag a+E_B b^\dag b$
is a free Hamiltonian of two uncoupled sites,
$V=a^\dag b+ab^\dag$
is their coupling, $H_h$ and $H_c$ are free Hamiltonians of the reservoirs, and
$H_{Ah}=(a+a^\dag)\otimes R_h$ and
$H_{Bc}=(b+b^\dag)\otimes R_c$
specify site-reservoir interactions. Here $a$ ($a^\dag$) and $b$ ($b^\dag$) are annihilation  (creation) operators for the sites A and B correspondingly. They satisfy the relations
\begin{eqnarray*}
aa^\dag+\delta a^\dag a&=1,\quad &aa+\delta aa=0,\\
bb^\dag+\delta b^\dag b&=1,\quad &bb+\delta bb=0,
\end{eqnarray*}
where $\delta=1$ for two-level systems and $\delta=-1$ for harmonic oscillators.

Our goal is to obtain a nontrivial expression for the steady-state heat flux from the hot reservoir based on the second-order perturbations in (\ref{EqPertEn})--(\ref{EqPertSt}) and to ascertain that it is positive, i.e. does not violate the second law of thermodynamics. 

If the sites are harmonic oscillators, there is no need for perturbation theory, because the Bogoliubov transformation \cite{BB} maps the system Hamiltonian $H_S=H_0+\nu V$ into a sum of two uncoupled oscillators (though each of them is coupled not to a single, but to both reservoirs). This considerably simplifies the task, and it is not so hard to derive and analyse the global master equation, which has been done in \cite{LevyKosloff}. The case of two-level systems is more intricate for the global approach, and our ``perturbed local approach'' can be useful. So, we restrict our analysis to this case ($\delta=1$).

We suppose that $E_A>E_B$. As shown in \cite{LevyKosloff}, this is a necessary condition for the violation of the second law of thermodynamics if we use the local approach. Also, the opposite case $E_A<E_B$ can be trivially mapped into this one, because, actually, we will use that $\beta_c>\beta_h$ only at the last moment; the case $E_A=E_B$ is less difficult and was analysed in \cite{Manrique} in the global approach. Moreover, for simplicity let $E_A-E_B=1$; the case of an arbitrary difference $E_A-E_B$ can be reduced to this one by a rescaling of the parameters of the system Hamiltonian. 

In the considered case, the eigenvalues and eigenvectors of the system Hamiltonian up to the second-order in $\nu$ terms are:
\begin{equation*}
\begin{split}
&\varepsilon_{00}=0, \quad \ket{e_{00}}=\ket{00},\quad \varepsilon_{11}=E_A+E_B, \quad \ket{e_{11}}=\ket{11},\\
&\varepsilon_{01}=E_B-\nu^2,\quad \ket{e_{01}}=\left(1-\frac{\nu^2}{2}\right)\ket{01}-\nu\ket{10},\\
&\varepsilon_{10}=E_A+\nu^2,\quad \ket{e_{10}}=\nu\ket{01}+\left(1-\frac{\nu^2}{2}\right)\ket{10}.
\end{split}
\end{equation*}
Here $\ket{00},\ket{01},\ket{10}$, and $\ket{11}$ denote the both sites in the ground states, only site $B$ in the excited state, only site $A$ in the excited state, and both sites in the excited states, correspondingly.

If we substitute these expressions to (\ref{EqMasterGlob})--(\ref{EqMasterDissip}), we will obtain the master equation
\begin{equation}\label{EqMaster}
\frac{\upd\rho}{\upd t}=-i[H_0+\nu V,\rho]+\lambda^2 \mathcal D_h(\rho)+\lambda^2 \mathcal D_c(\rho),
\end{equation}
where
\begin{equation}\label{EqDiss}
\begin{split}
\mathcal D_s(\rho)&=\sum_{j=A,B}\gamma_{js}
\left(F_{js}\rho F_{js}^\dag-\frac12\{F_{js}^\dag F_{js},\rho\}\right.\\
&\left.+\exp(-\beta_s\omega_j)\left(F_{js}^\dag\rho F_{js}-\frac12\{F_{js}F_{js}^\dag,\rho\}\right)\right),
\end{split}
\end{equation}
$s=h,c$. For shortness, we have adopted the notations $\gamma_{js}=\gamma_s(\omega_j)$ (where $\gamma_s$ is a correlation function of the corresponding reservoir), $F_{js}=F_s(\omega_j)$, where $F_h=a+a^\dag$, $F_c=b+b^\dag$,
\begin{equation*}
\begin{split}
\omega_A&=\varepsilon_{11}-\varepsilon_{01}=
\varepsilon_{10}-\varepsilon_{00}=E_A+\nu^2,\\
\omega_B&=\varepsilon_{11}-\varepsilon_{10}=
\varepsilon_{01}-\varepsilon_{00}=E_B-\nu^2.
\end{split}
\end{equation*}

It turns out that
\begin{equation*}
\begin{split}
F_{Ah}&=(1-\nu^2)a-
\nu(a^\dag a-aa^\dag)b,\\
F_{Bh}&=\nu^2a+
\nu(a^\dag a-aa^\dag)b,\\
F_{Ac}&=\nu^2b-
\nu a(b^\dag b-bb^\dag),\\
F_{Bc}&=(1-\nu^2)b+
\nu a(b^\dag b-bb^\dag).
\end{split}
\end{equation*}

Also relation (\ref{EqGamma}) has been used in (\ref{EqDiss}). Finally, we have neglected the Lamb shift Hamiltonian: it leads to modifications of the parameters of the system Hamiltonian ($E_A$, $E_B$ and the coefficient before $V$); this does not play a crucial role in our problem. Though, in other problems it can play a significant role \cite{Flem}.

The heat flux from the hot reservoir is given by the formula
\begin{equation}\label{EqFlux}
\mathcal J_h=\mathrm{Tr}[\mathcal D_h(\rho_*)(H_0+\nu V)],
\end{equation}
where $\rho_*$ is a steady-state density operator. The substitution of (\ref{EqDiss}) into (\ref{EqFlux}) yields the expression
\begin{equation}\label{EqFlux2}
\mathcal J_h=\lambda^2(J_h^{(1)}+J_h^{(a)}\langle a^\dag a\rangle+
J_h^{(b)}\langle b^\dag b\rangle+J_h^{(X)}\langle X\rangle),
\end{equation}
where
\begin{equation*}
\begin{split}
J_h^{(1)}= \gamma_{Ah}E_A \exp(-\beta_h \omega_A) &+ 
  \gamma_{Bh} E_B\nu^2\exp(-\beta_h \omega_B) \\&- 
 \gamma_{Ah}  E_B\nu^2 \exp(-\beta_h \omega_B)
 \end{split}
\end{equation*}
\begin{equation*}
\begin{split}
J_h^{(a)}=\gamma_{Ah} [1 + \exp(-\beta_h\omega_A)]
 [E_A\left(1 - \nu^2) - 
    E_B \nu^2\right)] 
\end{split}
\end{equation*}
\begin{equation*}
\begin{split}
J_h^{(b)}= \nu^2 (&\gamma_{Ah} E_A [1 + 
        \exp(-\beta_h\omega_A)] \\+ &\gamma_{Bh} E_B [1 + 
        \exp(-\beta_h\omega_B)])
\end{split}
\end{equation*}
$$
J_h^{(X)}=  \gamma_{Ah} \nu E_A[1 + \exp(-\beta_h \omega_A)],
$$
$X=a^\dag b+ab^\dag$ (in our case $V=X$), and $\langle T\rangle$ stands for $\mathrm{Tr}[\rho_*T]$ for an arbitrary operator $T$.

The averages $\langle a^\dag a\rangle$, $\langle b^\dag b\rangle$, $\langle X\rangle$, and $\langle Y\rangle=i\langle a^\dag b-ab^\dag\rangle$ satisfy the closed system of equations (follows from master equation (\ref{EqMaster})--(\ref{EqDiss}) rewritten in the Heisenberg representation)
\begin{equation}\label{EqSystem}
\begin{split}
\frac{\upd}{\upd t}\langle a^\dag a\rangle&=-\lambda^2M^{(aa)}\langle a^\dag a\rangle+\frac{\lambda^2\nu}2 M^{(aX)}\langle X\rangle\\&-\nu\langle Y\rangle+\lambda^2c^{(a)},\\
\frac{\upd}{\upd t}\langle b^\dag b\rangle&=-\lambda^2M^{(bb)}\langle b^\dag b\rangle+\frac{\lambda^2\nu}2 M^{(aX)}\langle X\rangle\\&+\nu\langle Y\rangle+\lambda^2c^{(b)},\\
\frac{\upd}{\upd t}\langle X\rangle&=\lambda^2\nu M^{(ax)}\langle a^\dag a\rangle+
\lambda^2\nu M^{(aX)}\langle b^\dag b\rangle\\&-  \frac{\lambda^2}2M^{(XX)}\langle X\rangle+\langle Y\rangle+2\lambda^2\nu c^{(X)},\\
\frac{\upd}{\upd t}\langle Y\rangle&=2\nu \langle a^\dag a\rangle-2\nu \langle b^\dag b\rangle\\&-\langle X\rangle-\frac{\lambda^2}2M^{(XX)}\langle Y\rangle,
\end{split}
\end{equation}
where
\begin{equation*}
\begin{split}
M^{(aa)}&=\gamma_{Ah}(1 - 2\nu^2) [1 + 
    \exp(-\beta_h\omega_A)] \\&+ \gamma_{Bc}\nu^2 [1 + 
     \exp(-\beta_c\omega_B)] \\&+ \gamma_{Ac}\nu^2[1 + \exp(-\beta_c\omega_A)],
\end{split}
\end{equation*}
\begin{equation*}
\begin{split}
M^{(bb)}&=\gamma_{Bc}(1 - 2\nu^2) [1 + 
    \exp(-\beta_c\omega_B)] \\&+ \gamma_{Bh}\nu^2 [1 + 
     \exp(-\beta_h\omega_B)] \\&+ \gamma_{Ah}\nu^2[1 + \exp(-\beta_h\omega_A)],
\end{split}
\end{equation*}
\begin{equation*}
\begin{split}
M^{(XX)}&=\gamma_{Bh}\nu^2 [1 + \exp(-\beta_h\omega_B)] \\&+ 
\gamma_{Ah} (1 - \nu^2) [1 + 
    \exp(-\beta_h\omega_A)] \\&+ \gamma_{Ac}\nu^2 [1 + 
     \exp(-\beta_c\omega_A)] \\&+ \gamma_{Bc} (1 - \nu^2) [1 + \exp(-\beta_c\omega_B)],
\end{split}
\end{equation*}
\begin{equation*}
M^{(aX)}=\gamma_{Bc} [1 + 
     \exp(-\beta_c\omega_B)] - \gamma_{Ah} [1 + 
     \exp(-\beta_h\omega_A)],
\end{equation*}
\begin{equation*}
\begin{split}
c^{(a)}=\gamma_{Ah}
  \exp(-\beta_h\omega_A)(1 - 2 \nu^2) &+ \gamma_{Bc}\nu^2\exp(-\beta_c\omega_B) \\&+ \gamma_{Ac}\nu^2\exp(-\beta_c\omega_A),
\end{split}
\end{equation*}
\begin{equation*}
\begin{split}
c^{(b)}=\gamma_{Bc}
  \exp(-\beta_c\omega_B)(1 - 2 \nu^2) &+ \gamma_{Bh}\nu^2\exp(-\beta_h\omega_B) \\&+ \gamma_{Ah}\nu^2\exp(-\beta_h\omega_A),
\end{split}
\end{equation*}
$$c^{(X)}=\gamma_{Ah}\exp(-\beta_h\omega_A) - 
 \gamma_{Bc}\exp(-\beta_c\omega_B).$$
 
Note that if we drop the terms of orders $\lambda^2\nu$ and $\lambda^2\nu^2$ in (\ref{EqMaster}) and (\ref{EqSystem}), we will obtain the corresponding equations from \cite{LevyKosloff}.
   
All left-hand sides in (\ref{EqSystem}) are zero since the averaging is performed over a steady state. So, we obtain a system of linear equations. The substitution of its solution into (\ref{EqFlux2}) yields the final result
\begin{equation}\label{EqFinal}
\begin{split}
\mathcal J_h=\lambda^2\nu^2&\left(\gamma_{Ac}E_A
\frac{1-\exp(-E_A\Delta\beta)}{1+\exp(E_A\beta_h)}\right.\\&\left.+
\gamma_{Bh}E_B
\frac{\exp(E_B\Delta\beta)-1}{\exp(E_B\beta_c)+1}\right)+o(\lambda^2\nu^2),
\end{split}
\end{equation}
$\Delta\beta=\beta_c-\beta_h$. Here we have dropped not only  terms of  order $O(\lambda^2\nu^3)$, but also terms of  order $O(\lambda^3)$, because the master equation itself is derived up to  order $\lambda^2$. If we, however, hold terms of order $O(\lambda^3)$, the expression will be much more cumbersome, but  this will not affect the sign of $\mathcal J_h$. Also, as we noticed at the end of the previous section, we assume that $\nu^2\ll\lambda$ (otherwise, the derivation of terms of order $\lambda^2\nu^2$ makes no sense without corrections to the master equation of order $\lambda^3$, which are greater in this case).

In the case of an arbitrary difference $\Delta E=E_A-E_B>0$, the resulting formula (\ref{EqFinal}) is still valid with $\nu$ substituted by $\nu/\Delta E$.

We see that, since $\Delta\beta>0$, the heat flux from the hot reservoir is always positive in accordance with the second law of thermodynamics.

\section{Remark}
We assumed that the perturbation $\nu V$ in (\ref{EqHs}) does not change the multiplicities of the eigenenergies. The opposite case is more tricky. If there is a removal of degeneracy, the distances between the new energy levels are small (formally, they are infinitesimal as $\nu\to0$) and, hence, may be less than or comparable to the line broadening caused by the interaction with the reservoir. In this case, the quantum master equation approach or at least the secular approximation used in the microscopic derivation of the  master equation is, generally speaking, not valid \cite{MayKuhn,AccLuVol,BP}. So, equation (\ref{EqMasterGlob}), which we are based on, may be inadequate in this case.

A possible solution is to go one step back from the master equation in the Gorini--Kossakowski--Sudarshan--Lindblad \cite{GKS,Lindblad} (GKSL) form (\ref{EqMasterGlob}) to the more general Redfield equation \cite{MayKuhn,BP,Redfield} and not to apply the secular approximation for pairs of Bohr frequencies with small differences. The resulting master equation will be not of the GKSL form, but it will reproduce the dynamics more adequately (the adequacy of the secular approximation is discussed in \cite{Flem,Plenio}). It is interesting to mention that, in \cite{Breuer}, the local approach was introduced exactly to avoid the secular approximation due to small energy differences, but, nevertheless, to stay in the equation of the GKSL form.

However, two extreme cases can be discriminated. If the line broadening is much smaller than the distances between the levels from this new multiplet ($\lambda^2\ll\nu\ll1$), we can still use the quantum matrix equation in form (\ref{EqMasterGlob}), i.e., with the full secular approximation. But if we restrict ourselves to the zeroth-order expansion of eigenenergies and eigenstates (\ref{EqPertEn})--(\ref{EqPertSt}) in the master equation (local approach), we should ignore the degeneracy of zeroth-order levels.

On the contrary, if the line broadening is much larger than the width of this multiplet (difference between the maximal and minimal levels of it), i.e. $\nu\ll\lambda^2\ll1$, the removal of degeneracy should be ignored, and the full secular approximation can be applied. 

So, in both cases, we stay in an equation of the GKSL form (\ref{EqMasterGlob}).

\section{Conclusions} The microscopically derived global master equation for quantum  dynamics in open networks can be too difficult to construct and/or to analyse. The local master equation simplifies this task, but its range of validity can be too restrictive. Here we have proposed an intermediate approach, which uses  stationary perturbation theory to obtain corrections to the local approach up to a desirable order of the inter-site coupling constant. This technique bridges the local and global approaches to construction of quantum master equations.

\acknowledgments
The authors are grateful to Mikhail Ivanov, Arseniy Mironov, and Sergey Kozyrev for helpful discussions. This work is supported by the Russian Science Foundation under grant 14-50-00005.

\end{document}